%Paper: astro-ph/9511136
%From: Leszek_Zdunik@camk.edu.pl (Leszek Zdunik)
%Date: Tue, 28 Nov 1995 13:03:44 +0100

\def\vm{\zeta_{\rm m}}
\def\vd{\zeta_{\rm d}}
\def\vmean{\overline\zeta}
\def\etamean{\overline\eta}
\def\rhomean{\overline\rho}
\def\tv{\tau_\zeta}
\def\Cs{C_\eta}
\def\Cb{C_\zeta}
\def\Cbm{C_{\zeta_{\rm m}}}
\def\Cbd{C_{\zeta_{\rm d}}}
\def\qdm{q_{\rm dm}}
\def\rs{\rho_{15}}
\def\os{\omega_4}
\def\xs{x_{01}}
\def\vunit{{\rm g\,cm^{-1}\,s^{-1}}}
\def\rhou{{\rm g\,cm^{-3}}}
\def\epstil{\tilde\epsilon}
\def\Ocm{{\Omega_{\rm c}}_{\min}}
\def\Ok{\Omega_{\rm Kepler}}
\def\Tcr{T_{\rm crit}}
\def\dotsa{\hbox to 2cm{\dotfill}}

\font\titlef=cmbx10 scaled \magstep3
\font\authorf=cmssbx10 scaled \magstep2
\font\addressf=cmss10 scaled \magstep1
\def\Title#1{\centerline{\titlef #1}\vskip20pt plus 5pt}
\def\Author#1{\centerline{\authorf #1}\vskip10pt plus 5pt}
\def\Address#1{\centerline{\addressf #1}}

\long\def\Abstract#1{\vskip10pt\noindent\hfill\vbox{\hsize=0.9\hsize
\noindent{\bf Abstract.} #1}\hfill}
\newdimen\stdparindent\stdparindent\parindent

\def\ref{\par
   \hangindent\stdparindent\hangafter1\noindent\ignorespaces}

\font\titleaf=cmbx10 scaled \magstep1
\font\titlebf=cmbxsl10
\def\titlea#1{\vskip10pt plus 5ptminus 5pt {\noindent\titleaf #1}}
\def\titleb#1{\vskip5pt plus 2pt minus 2pt {\noindent\titlebf #1}
\vskip 5pt minus2pt}

\rightline{NCAC preprint No. 306~~~~~November 1995}
\vskip2truecm

\Title{Damping of GRR instability by direct URCA reactions}
\Author{J.L. Zdunik}
\Address{N. Copernicus Astronomical Center}
\Address{ Bartycka 18, 00-716 Warszawa, Poland}
\Address{  e-mail: jlz@camk.edu.pl}
\vskip10pt
\centerline{Accepted for publication in {\it Astronomy and Astrophysics}}

\Abstract{
The role of direct URCA reactions in damping of the gravitational
radiation driven instability is discussed. The temperature at which
bulk viscosity suppresses completely this instability is calculated.
\hfill\break\indent
The results are obtained analytically using recent calculations
performed in the case of bulk viscosity due to the modified URCA
processes (Lindblom 1995; Yoshida \& Eriguchi 1995).\hfill\break\indent
The bulk viscosity caused by direct URCA reactions is found to reduce
significantly
the region of temperatures and rotation frequencies where a
neutron star is subject to GRR instability.
}

\titlea{Introduction}

An absolute upper bound on angular velocity of rotating star is
given by a shedding limit, ie. the Keplerian frequency $\Ok$,
defined as a velocity of
a test particle on a circular orbit on equator.

However it is possible that a star rotating with $\Omega<\Ok$ is
unstable against some perturbations. In this case this
instability sets the upper limit on star's rotation.

Neutron stars rotating with very high angular velocity could be
subject to secular instability caused by the gravitational
radiation reaction (GRR). This possibility has been pointed out
by Chandrasekhar (1970) and then
shown to be generic feature of rotating star (Bardeen at al.
1977; Friedman \& Schutz 1978; Friedman 1978).

Further studies have shown that dissipation processes, such as
bulk and shear viscosity, moderate this instability and in
some ranges of temperature damp it completely. For
temperatures lower than $10^7$~K shear viscosity ($\eta \sim T^{-2}$)
stabilize stars rotating with $\Omega<\Ok$ (Cutler \& Lindblom
1987; Ipser \& Lindblom 1991). At high
temperatures $T>10^{10}$~K  the gravitational radiation driven
instability is damped by bulk
viscosity resulting from modified URCA processes (Cutler at al. 1990;
Ipser \& Lindblom 1991).

It was recently pointed out by Lattimer et al. (1991) that the
proton fraction inside neutron stars could be sufficiently large
to make
direct URCA processes possible. Haensel \& Schaeffer (1992) calculated
the bulk viscosity of hot neutron--star matter caused by direct
URCA reactions. In the interesting range of temperatures
this viscosity appears to be many orders of magnitude
larger than bulk viscosity corresponding to modified URCA processes.
Thus damping of GRR instability should be much more effective.

In present paper we analyse the consequences of direct URCA
reactions for the instability driven by gravitational radiation
reaction. We use the results of previous calculations performed
by Lindblom (1995) and Yoshida \& Eriguchi (1995) for bulk
viscosity due to the modified URCA processes. Our conclusions are
obtained analytically using approximate formulae describing the
damping of GRR instability for given value of viscosity.
However, because of the strong dependence of shear and bulk
viscosities on temperature, it seems that our estimates are
very accurate.

\goodbreak
\titlea{Calculations and Results}

\titleb{Bulk viscosity from direct and modified URCA processes.}

The bulk viscosity of the matter is given by the formula:
$$\zeta = -{\lambda\cdot C^2\over \omega^2 + 4\lambda^2 B^2/n^2}
\eqno(1)$$
(Sawyer 1989) where $n$ is the total baryon number density and
coefficient $\lambda$ describes the linear response of the
rates of the beta and inverse--beta reactions to the departure
from beta equilibrium (given by the value
of $\delta\mu=\mu_{\rm n}-\mu_{\rm p}-\mu_{\rm e}$).
The coefficients $B$ and $C$ determine the change of $\delta\mu$
due to the change in the chemical composition, at fixed density of
the matter ($B$) and due to the compression or
rarefaction of the matter, where the weak interactions are kept
frozen ($C$).

In our considerations we calculate bulk viscosity using
``high--frequency'' approximations both
in the case of direct and modified URCA reactions.
This approximations is valid if the second term in the
denominator of Eq.(1) is negligibly small compared to
$\omega^2$. The angular frequency of the pulsational modes
which determine the maximum angular velocity of the star
is of the order of $10^4$~s$^{-1}$.
In our case we can estimate the value of $(2\lambda B/n \omega)^2$
as:
$\sim 0.05 \cdot T_{\rm MeV}^8/\os^2$ for direct URCA reactions
and $\sim 10^{-9} \cdot T_{\rm MeV}^{12}/\os^2$ for modified URCA
processes ($T_{\rm MeV}=kT/1$~MeV and
$\os=\omega/10^4$~s$^{-1}$).
We see that at, say, $T<10^{10}$~K we can safely use
``high--frequency'' approximation ($(2\lambda B/n \omega)^2\ll 1$).

In this limit bulk viscosity caused by modified URCA processes
is given by:

$$\vm = 6.0 \cdot 10^{17} \left({\rs\over\os}\right)^2 {T_9}^6~~~\vunit,
\eqno(2)$$

\noindent where $\rs=\rho/10^{15}~\rhou$ and $T_9=T/10^9$~K.

The above formula was used by Ipser \& Lindblom (1991) and had been
derived by Sawyer (1989) but published originally with a
typographical error (Lindblom 1995).

The bulk viscosity due to the direct URCA processes is given in
the ``high--frequency'' limit by
(Haensel \& Schaeffer 1992):

$$\vd = 8.9 \cdot 10^{24} {m_{\rm n}^*\over{m_{\rm n}}}
{m_{\rm p}^*\over{m_{\rm p}}}
\left( x {n\over n_0}\right)^{1/3} {C^2_{100}\over \os^{2}}
{T_9}^4~~~\vunit,\eqno(3)$$

\noindent where $m_{\rm n}$, $m^*_{\rm n}$, $m_{\rm p}$, $m^*_{\rm p}$ are
neutron and proton true
and effective masses respectively, $n_0$ --- saturation density of
nuclear matter ($n_0=0.16$~fm$^{-3}$), $x=n_{\rm p}/n$ --- proton fraction,
and\ ($C_{100}$ --- $C$ in units 100~MeV). The parameter
$C_{100}$ is density (and model) dependent and in the region of
densities interesting for our problem is of the order of one
(between 1 and 2, see Fig. 2 in the paper by Haensel \&
Schaeffer (1992)).
It should be mentioned that to derive bulk viscosity of neutron--star matter
Sawyer (1989) used free Fermi--gas ($n$--$p$--$e$) approximation and Haensel
\& Shaeffer (1992) take into account the effect of nucleon--nucleon
interaction on the nuclear symmetry energy of the matter.
These effects play a crucial role in determining the proton fraction $x$
(sufficiently large value of $x$ allows direct URCA reactions to take place)
and also influence the function $C(\rho)$.
We disscus here
two models corresponding to two different depencences of the
nuclear symmetry energy on density (Model I and II of Haensel \&
Schaeffer (1992)).

We will use a simplified version of formula (3)

$$\vd = 6.5 \cdot 10^{24} q \xs^{1/3} {\rs^{1/3}\over\os^2} {T_9}^4~~~\vunit,
\eqno(4)$$

\noindent where $\xs=x/0.1$ and we have introduced parameter
$q={m_{\rm n}^*\over{m_{\rm n}}} {m_{\rm p}^*\over{m_{\rm p}}}
C^2_{100}$, which is dimensionless and of the order of one.

There are some physical limits on the region of densities and
temperatures where direct URCA processes are operating and above
formula is applicable.

First, the value of the proton fraction $x$ should be
sufficiently large ($x>x_{\rm crit}\approx{1\over9}$). This condition
defines minimum density below which we have to do with modified
URCA processes. This minimum density $\rho_{\min}$ was calculated
by Haensel and Schaeffer (1992) and is equal to $0.53\cdot 10^{15}~\rhou$
and $0.72\cdot 10^{15}~\rhou$ for their models II and I respectively.
Thus direct URCA processes are operating in a central regions of
a neutron stars.

Second, there exist a critical temperature $\Tcr$ below which
neutrons become superfluid and protons superconductive.
In this region reaction rates of the URCA processes (both direct
and modified) are significantly reduced.
The estimation of $\Tcr$ gives the value of the order $10^9$~K
(Page 1994).

\titleb{Damping of GRR instability by ``direct URCA'' bulk viscosity.}

To calculate the effect of bulk viscosity due to the direct URCA
reactions on the damping of GRR instability we use previous
results published by Lindblom (1995) and Yoshida \& Eriguchi
(1995)
obtained taking into account modified URCA processes only.

We restrict ourself to the Newtonian stellar model of the mass
$1.5\, M_\odot$  built of polytropic
matter with index $n=1$ (pressure $P \sim \rho^2$).

{}From physical point of view the mechanism of the damping of the
oscilations which tends to grow as a reaction to gravitational
radiation is in both
cases (modified or direct URCA) the same. Namely the source of
damping is departure from beta equilibrium due to compression or
rarefaction of the matter during oscillations.
The only difference are reactions between constituents of the matter
which lead to above nonequilibrium processes.

The conclusion is that for a given value of $\vd$ the effect on
damping of GRR instability is the same as for the same value of $\vm$.

Above statement is true if we consider a small piece of matter
in a star and would be true if we have constant viscosity
$\zeta$ throghout a star (or the dependence of $\vm$ and $\vd$
on $\rho$ is the same). But this is not the case and damping of
GRR instability by viscosity is a global feature of a star. The
functional dependence $\zeta(\rho)$ enters ours considerations
through the integral defining the damping time due to the bulk
viscosity of matter (Ipser \& Lindblom 1991):

$${1\over\tv}={1\over 2E}\int\zeta\,\delta\sigma\,\delta\sigma^*\,d^3x,
\eqno(5)$$

\noindent where $\sigma=\nabla_a v^a$ is the
expansion ($v^a$ --- fluid velocity).

We take this fact into account defining the parameter $\Cb$:

$$\Cb={1\over 2E}\int{\,\zeta\,\over\,\vmean\,}\,\delta\sigma\,\delta\sigma^*\,
d^3x,\eqno(6)$$

\noindent where $\vmean=\zeta(\rhomean)$ and $\rhomean$ is the
mean density of a star. It should be stressed that the value of
$\Cb$ depends on the pulsation mode.

Using this definition we can rewrite Eq. (5) as

$${1\over\tv}=\Cb\cdot\vmean\eqno(7)$$

Assuming, that for the same value of $\tv$ GRR instability is
damped
to the same extent
we can very easily estimate the
the temperature $T_{\rm d}$ at which bulk viscosity caused by direct URCA
reactions completely damps this instability
($\Omega_{\max}=\Ok$) having corresponding value $T_{\rm m}$ for
modified URCA. The resulting formula is:

$$\eqalign{\log T_{\rm d} = &1.5\log T_{\rm m} - 6.26 +
0.25\log\left({\Cbm\over\Cbd}\right)\cr
&+0.25\log{\rs^{5/3}\over q \xs^{1/3}}}\eqno(8)$$

For modified URCA this temperature is equal $T_{\rm
m}=10^{10}$~K (Lindblom 1995; Yoshida \& Eriguchi 1995).
Thus in the case of direct URCA and polytropic star with index
$n=1$ and mass $M=1.5 M_\odot$
considered by Lindblom ($\rhomean_{15}=0.36$) we obtain:

$$\log T_{\rm d} = 8.55 - 0.25\log(\qdm) - 0.25\log{q\xs^{1/3}}\eqno(9)$$

\noindent where we have introduced the parameter
$\qdm={\Cbd/\Cbm}$.

This parameter measures the relative difference between damping
by modified and direct URCA reactions resulting from a different shape
of $\vm$ and $\vd$ throughout the star.

To estimate the value of $\qdm$
we should evaluate two integrals:
%in two lines
$$\eqalign{
\Cbm&={1\over 2E}\int{\vm\over\vmean_{\rm m}}\,\delta\sigma\,\delta\sigma^*
\,d^3x,\cr
\Cbd&={1\over 2E}\int{\vd\over\vmean_{\rm d}}\,\delta\sigma\,\delta\sigma^*
\,d^3x,\cr}\eqno(10)$$

A more careful analysis of the radial dependence of integrands
with the help of the Fig. 2 in the paper by Ipser \& Lindblom
(1991) leads to the following approximation of $\qdm$:

$$\qdm\approx {\int \phi_{\rm d} d^3x \over\int \phi_{\rm m} d^3x},\eqno(11)$$

\noindent where $\phi_{\rm d}$ and $\phi_{\rm m}$ are integrands in $\Cbd$
and $\Cbm$ normalised as in Ipser \& Lindblom (1991) (the maximum
value of each integrand is one).

The crucial point in estimating of $\Cbd$ is the radius $r_{\rm d}$ of the
central sphere in the star where direct URCA processes are operating.
Stricly speaking in calculation of $\Cbd$ we should restrict
ourself just to this central region:
$$\Cbd={1\over 2E}\int_0^{r_{\rm d}} {\vd\over\vmean_{\rm d}}\,\delta\sigma\,
\delta\sigma^*\,d^3x,\eqno(12)$$
\noindent or, in other words, put $\vd=0$ for $r>r_{\rm d}$.

We can determine the value of $r_{\rm d}$ for the considered model of
a star ($n=1$ polytrope, $\rho_{\rm central}=1.2\cdot10^{15}~\rhou$).
For models I and II of Haensel \& Schaeffer (1992) we obtain:

Model I\phantom{I}~~~~$\rho_{\min}=0.72\cdot 10^{15}~\rhou$~~~~
$r_{dI\phantom{I}}=0.53\cdot R$

Model II~~~~$\rho_{\min}=0.53\cdot 10^{15}~\rhou$~~~~
$r_{dII}=0.65\cdot R$

\noindent where $R$ is the radius of the star, $R=12.53$~km.

Having the value of $r_{\rm d}$ and comparing it with the Fig 2.
of Ipser \& Lindblom (1991) we could answer the question whether
direct URCA processes are operating in the region where bulk
viscosity damps effectively GRR instability .

These two effects limit the region of a star where damping of
GRR instability by direct URCA takes place.
The oscillatory modes which are subject to this instability reveal
significant changes in density in the outer region of a star.
{}From the Fig.2 in the paper by Ipser \& Lindblom (1991)
we could see that function $\delta\sigma\,
\delta\sigma^*$ starts growing at $r_{\sigma}\approx 0.4\cdot R$.

Thus the main contribution to the integral $\Cbd$ comes from the
region between $r_{\sigma}$ and $r_{\rm d}$.

This considerations enables us to estimate the value of $\qdm$:
$\qdm \sim 1/5 \div 1/10$ for model II and $\qdm \sim 1/20 \div 1/100$
for model I.

Taking into account these estimations we can rewrite Eq.(9) as

$$\eqalign{\log T_{\rm d} &\approx 8.9-0.25\log{q\xs^{1/3}}~~~~~~~{\rm
Model~I}\cr
\log T_{\rm d} &\approx 8.7-0.25\log{q\xs^{1/3}}~~~~~~~{\rm
Model~II}\cr}\eqno(13)$$

We see that even though
damping of GRR instability by direct URCA reactions is limited
to rather narrow region of the star the process is effective for
temperatures larger than $10^9$~K.

\titleb{Critical angular velocities of rotating NS.}

Using some approximations we can estimate not only the minimum
temperature $T_{\rm d}$ at which bulk viscosity caused by direct URCA
processes completely damps GRR
instability but also the function $\Omega_{\rm c}(T)$ in the region
where star is subject to this instability for sufficiently large
$\Omega$ ($\Omega_{\rm c}<\Omega<\Ok$).

For ``modified URCA bulk viscosity'' $\Omega_{\rm c}(T)$ was determined
by Lindblom (1995) and Yoshida \& Eriguchi (1995).
In this case GRR instability is completely
damped for temperatures lower than $T_\eta=10^7$~K by shear
viscosity and for temperatures greater than $T_{\rm m}=10^{10}$~K by
bulk viscosity. Thus for $T<T_\eta$ and $T>T_{\rm m}$ we have $\Omega_{\rm
c}(T)=\Ok$.
Between $T_\eta$ and $T_{\rm m}$ function
$\Omega_{\rm c}(T)$ is smaller than $\Ok$ having minimum
equal to $\approx 0.95 \Ok$ at $T_{\min}\approx
2\cdot 10^9$~K (Fig. 1 of Lindblom (1995)).

In the case of the direct URCA reactions $\Omega_{\rm c}(T)$ differs
from that of Lindblom (1995) in the region
which corresponds to the damping due to bulk viscosity.
This part of the function $\Omega_{\rm c}(T)$ is increasing and reaches its
constant and maximum value $\Ok$ at temperature
$T=T_{\rm d}$, significantly lower than $T_{\rm m}$. Furthermore the slope
of $\Omega_{\rm c}(T)_{\rm d}$ is smaller than for modified URCA reactions
$\Omega_{\rm c}(T)_{\rm m}$. The reason is simply the different power in
temperature dependence of $\vd(T)$ and $\vm(T)$.
{}From formula (8) we get:
$$\left({d\Omega_{\rm c}\over d \log T}\right)_{\rm d}(T_2)=
{2\over3}\left({d\Omega_{\rm c}\over d \log T}\right)_{\rm m}(T_1),
\eqno(14)$$
\noindent and two derivatives are calculated for the same value
of $\Omega_{\rm c}$ (or for ${\tv}_{\rm d}(T_2)={\tv}_{\rm m}(T_1)$).

Because of the rather strong dependence of $\zeta$ and  $\eta$
on temperature
(${\zeta/\eta}\sim T^6$)
the region where
shear and bulk viscosities play significant and comparable role
is very narrow in temperature and so minimum of the function
$\Omega_{\rm c}(T)$ is quite well defined by intersection of
the parts of these dependences calculated for shear viscosity
and bulk viscosity separately.

More precisely we can determine the temperature $T_{\min}$ at
which $\Omega_{\rm c}(T)$ has its minimum analysing the function:
$${1\over\tau_\eta(\Omega)}+{1\over\tau_\zeta(\Omega)}=
\beta(\Omega)\cdot \left({1\over\tau_\eta(0)}+
\epstil(\Omega)
\cdot {1\over\tau_\zeta(0)}\right)\eqno(15)$$
\noindent where dimensionless functions $\beta(\Omega)$ and
$\epstil(\Omega)$
were defined by Lindblom (1995) and Ipser \& Lindblom (1991)
and presented on graphs in
their paper.
Introducing, as in the case of bulk viscosity, the parameter $\Cs$:
$${1\over\tau_\eta}=\Cs\cdot\etamean\eqno(16)$$
\noindent we write the Eq.(15) in the form:
$${1\over\tau_\eta(\Omega)}+{1\over\tau_\zeta(\Omega)}=
\beta(\Omega)\cdot (\Cs\etamean+
\epstil(\Omega)
\cdot \Cb\vmean).\eqno(17)$$
The estimates of parameters $\Cs$ and $\Cb$ from Table 1 of
Lindblom (1995) gives the ratio $\gamma\equiv{\Cb/\Cs}$ to be of the
order 0.01 ($\gamma$ depends on the mode and is equal:
$0.67\cdot 10^{-2}$, $0.9\cdot 10^{-2}$, $1.2\cdot 10^{-2}$ for modes
$m=l=4,3,2$ respectively).
The difference between $\Cs$ and $\Cb$ reflects the fact that
shear and bulk viscosities damps GRR instability in two
different ways. For the same value of viscosity shear viscosity
damps this instability about two orders of magnitude more
effectively than bulk viscosity.

We can make a remark that to estimate
roughly the temperature where shear and bulk
viscosity play comparable role one should find the crossover point of
the functions
$\eta(T)$ and $\gamma\cdot\zeta(T)$ rather than $\eta(T)$
and $\zeta(T)$.

We see that to find the value of $T_{\min}$ we have to  analyse the
function:
$$\beta(\Omega)\cdot (\etamean+
\epstil(\Omega)
\cdot \gamma\vmean).\eqno(18)$$

This approach leads to the following formulae:

$$\eqalign{\log T_{\min} &\approx 8.3-{1\over6}\log{q\xs^{1/3}}~~~~~~~{\rm
Model~I}\cr
\log T_{\min} &\approx 8.2-{1\over6}\log{q\xs^{1/3}}~~~~~~~{\rm
Model~II}\cr}\eqno(19)$$

To obtain the above result we have extracted the value of
$\epstil(\Omega)$ from Fig. 13 of Ipser \& Lindblom (1991).

The corresponding values of $\Ocm$ could be estimated from
Fig.~1 of Lindblom (1995):
$$\eqalign{\Ocm &= 0.97\phantom{5}\,\Ok~~~~~~~{\rm Model~I}\cr
\Ocm &= 0.975\,\Ok~~~~~~~{\rm Model~II}\cr}\eqno(20)$$

The above results seems to be accurate. This approximate method applied
to bulk viscosity caused by modified URCA processes reproduces
very well the value $T_{\min}$
obtained by Lindblom (1995).

We conclude that if bulk viscosity due to the direct URCA
processes is operating at temperatures greater than $\sim 10^8$~K the
region in the $T-\Omega$ plane where a star is unstable with
respect to GRR instability is significantly smaller than in the
case of modified URCA processes.

\titlea{Conclusions}

Our analysis shows a significant role of direct URCA processes
in damping instability driven by gravitational radiation of a
rapidly rotating neutron stars.
In the absence of other restrictions (superfluidity of neutrons,
superconductivity of protons) ``direct URCA'' bulk viscosity is
effective at temperatures larger than $\sim 2\cdot 10^8$~K and
completely damps GRR instability at $\sim 5\cdot 10^8 \div 10^9$~K.
This process together with a lower bound defined by shear
viscosity limits the allowed region where a rapidly rotating
star could be unstable with respect to emission of gravitational
waves to temperatures from the interval $10^7 \div 10^9$~K. To
be subject to this instability a star should rotate with high
angular velocity which appears to be very close to the
Keplerian one --- at most $2.5\div3\,\%$ lower than the
shedding limit.

Taking into account the posibbility of superfluidity of neutrons
and superconductivity of protons below critical temperature
$\Tcr$ would slightly change our conclusions.
Theoretical calculations of $\Tcr$ lead to a very uncertain and
model dependent results (for a discussion see e.g., Page 1994).
It seems that in the density range where direct URCA reactions
are possible ($\rho>\rho_{\min}$) the critical temperature is
determined by neutron $^3{\rm P}_2$ pairing and lies between $10^8$~K
and $6\cdot 10^9$~K for various theoretical models (Page 1994,
and references therein).
Superfluidity of neutrons
and superconductivity of protons reduce URCA reaction rates,
although it is not obvious whether bulk viscosity is switched off
exponentially for $T<\Tcr$ ($\sim \exp(-\Tcr/T)$) or
smoothly in a rather broad temperature
region near $\Tcr$.
Recent calculations performed by Haensel at al. (1995)
support latter possibility and it seems that ``direct URCA viscosity''
could play a significant role even at temperatures few times lower than $\Tcr$
(Haensel 1995).

It should be mentioned that in the region where neutrons form a superfluid
(at $T < \Tcr$) the shear viscosity of the matter
results from electron--electron scattering and is of the same
order as in the case of neutron--neutron collisions (Ipser \&
Lindblom 1991).
Thus superfluidity of neutrons would change the shear viscosity
much less than the bulk one and
the net effect would be shifting of the temperature at which
shear and bulk viscosity play comparable role to a higher value
and lowering of the corresponding $\Ocm$.

Although superfluidity of the matter supresses the role of viscosity in
damping of GRR instability it also generates new dissipative mechanisms
(for review see e.g. Pines \& Alpar 1992).
Recently Lindblom \& Mendell (1995) using simple analytical model
estimated the role of so called ``mutual friction'',
the most important effect in damping of GRR instability.
They concluded that this dissipative effect completely suppresses
the GRR instability below $\Tcr$.

The next point which could influence our result is a
size of the central region in a star where direct URCA
processes are operative. On the contrary significant density changes in
modes which becomes unstable
with respect to gravitational radiation reaction are limited to
the outer region of a star. Thus there exist a shell where direct
URCA reactions could damp GRR instability. The thickness of this
shell depends on the density profile througout a star (and so on
equation of state) and also on the mass of the star.
Our calculations has been performed for $n=1$ polytropic star of
a mass $M=1.5\,M_\odot$.
For less
massive stars density in the ``damping region'' could be so small
that interesting shell would be very thin or even disappears.

The analysis of the stability of a rotating star in the
$T-\Omega$ plane should take into account cooling timescale of
this star. Strictly speaking a star becomes unstable with respect
to GRR instability when the growing time of this instability is
short compared to the timescale at which a star would cool
enough to leave the ``instability region'' (Yoshida \& Eriguchi 1995).
The cooling
timescale is many orders of magnitude smaller for direct URCA
processes than for the modified ones. Thus in the newly born
neutron stars the interesting region will cool to $10^9$~K in
minutes and to $T_{\min} \sim 10^8$~K in days (Haensel \&
Schaeffer 1992).
Yoshida \& Eriguchi (1995) published functions
$\Omega_{\max}(T)$ for several values of the growing timescale of
GRR instability.
Their results show that this effect
would shift $\Ocm$ to a little higher value.

\bigskip
\noindent{\it Acknowledgements.} I would like to thank P. Haensel for helpful
comments and
discussions.
This work was supported in part by KBN Grant No. 2P30401407.

\vskip10pt
{\noindent\bf References}
\vskip10pt minus 5pt
\ref Bardeen J.M., Friedman J.L., Schutz B.F., Sorkin R., 1977,
ApJ 217, L49
\ref Chandrasekhar S, 1970, Phys. Rev. Letters 24, 611
\ref Cutler C, Lindblom L., 1987, ApJ 314, 234
\ref Cutler C, Lindblom L., Splinter R.J., 1990, ApJ 363, 603
\ref Friedman J.L., 1978, Comm. Math. Phys. 62, 247
\ref Friedman J.L., Schutz B.F., 1978, ApJ 222, 281
\ref Haensel P., 1995, private communication
\ref Haensel P., Schaeffer R., 1992, Phys. Rev. D 45, 4708
\ref Haensel P., Levenfish K.P., Yakovlev D.G., 1995, in preparation
\ref Ipser J.R., Lindblom L., 1991, ApJ 373, 213
\ref Lattimer J.M., Pethick C.J., Prakash M., Haensel P., 1991,
     Phys. Rev. Letters 66, 2701
\ref Lindblom L., 1995, ApJ 438, 265
\ref Lindblom L., Mendell G., 1995, ApJ 444, 804
\ref Page D., 1994, ApJ 428, 250
\ref Pines D., Alpar M.A., 1992, In: Pines D., Tamagaki R., Tsuruta S. (eds.)
The Structure and Evolution of Neutron Stars, Addison--Wesley, New York,p. 7
\ref Sawyer R.F., 1989, Phys. Rev. D, 39, 3804
\ref Yoshida S., Eriguchi Y., 1995, ApJ 438, 830

\bye